\documentclass[11pt,a4paper]{article}
\usepackage{cite,amssymb}
\newcommand{\be}{\begin{eqnarray}}
\newcommand{\ee}{\end{eqnarray}}
\newcommand{\bez}{\begin{eqnarray*}}
\newcommand{\eez}{\end{eqnarray*}}
\newcommand{\pa}{\partial}
\newcommand{\la}{\lambda}
\title{\bf Extension of Noncommutative Soliton Hierarchies}

\date{}

\author{Aristophanes Dimakis \\
 Department of Financial and Management Engineering, \\
 University of the Aegean, 31 Fostini Str., GR-82100 Chios, Greece \\
 dimakis@aegean.gr
 \and
 Folkert M\"uller-Hoissen \\ Max-Planck-Institut f\"ur Str\"omungsforschung \\
 Bunsenstrasse 10, D-37073 G\"ottingen, Germany \\
 fmuelle@gwdg.de }

\begin{document}

\renewcommand{\theequation} {\arabic{section}.\arabic{equation}}

\maketitle

\begin{abstract}
A linear system, which generates a Moyal-deformed two-dimensional
soliton equation as integrability condition, can be extended to a
three-dimensional linear system, treating the deformation parameter
as an additional coordinate.
The supplementary integrability conditions result in a first order
differential equation with respect to the deformation parameter,
the flow of which commutes with the flow of the deformed soliton equation.
In this way, a deformed soliton hierarchy can be extended to a
bigger hierarchy by including the corresponding deformation equations.
We prove the extended hierarchy properties for the deformed AKNS hierarchy,
and specialize to the cases of deformed NLS, KdV and mKdV hierarchies.
Corresponding results are also obtained for the deformed KP hierarchy.
A deformation equation determines a kind of Seiberg-Witten map from classical
solutions to solutions of the respective `noncommutative' deformed equation.
\end{abstract}

\section{Introduction}
\setcounter{equation}{0}
Because of their appearance (see Refs.~\citen{LPS01a,LPS01b}, in particular)
in certain limits of string theories (see Ref.~\citen{Doug+Nekr01} for a review),
noncommutative generalizations of soliton equations, obtained from classical models
essentially by replacing the ordinary commutative product of functions with a
noncommutative Moyal $\ast$-product, attracted considerable interest recently
\cite{Taka01,DMH00ncKdV,DMH00SW,Lega00,Lech+Popo01c,DMH01ncNLS,DMH01FK,Pani01,Gris+Pena03,
Cabr+Mori02,Mart+Pash03,Hama03a,Hama03b,Hama+Toda03a,Hama+Toda03b,Hama+Toda03c,
Nish+Rajp03,Zuev03}. One can say that such equations live on a `noncommutative
space', they are not simply ordinary evolution equations of non-commuting objects
such as matrices. We will use `noncommutative' (nc) in this sense.
\vskip.1cm

In previous work \cite{DMH00ncKdV,DMH01ncNLS} on noncommutative versions of certain
soliton equations, we were able to find a differential equation of first order
in the deformation parameter $\theta$, which allowed us to calculate solutions of
the noncommutative soliton equation order by order in $\theta$ from solutions
of the corresponding classical soliton equation. Thus, in these cases there is
a map from solutions of the `commutative' to solutions of the `noncommutative'
equation which is analogous to the {\em Seiberg-Witten} (SW) {\em map}
of Ref.~\citen{Seib+Witt99}. Because of this reason we call such a $\theta$-evolution
equation a \emph{SW deformation equation}.
\vskip.1cm

A systematic way of generating such an equation will be described now.
We consider a nonlinear equation (or a system of equations) which is integrable
in the sense that it can be expressed as integrability condition of a linear
system of equations. Prominent examples
are given by soliton equations (see \cite{Fadd+Takh87}, for example).
Let us look at a linear system of the form
\be
   \pa_x \psi  = {\cal U} \ast \psi \, , \qquad
   \pa_t \psi = {\cal V} \ast \psi
\ee
where ${\cal U},{\cal V}$ are $N \times N$ matrices of functions
(or suitable operators). The associative noncommutative $\ast$-product is defined by
\be
   f \ast g = \mathbf{m} \circ e^{\theta P/2} (f \otimes g) \, ,
  \qquad
   P = \pa_t \otimes \pa_x - \pa_x \otimes \pa_t     \label{Moyal}
\ee
using $\mathbf{m}(f \otimes g) = f g$, for functions $f,g$ of the coordinates $x,t$
and the deformation parameter $\theta$.
The integrability condition of the above system is
\be
   \pa_t {\cal U} - \pa_x {\cal V} + [{\cal U},{\cal V}]_\ast = 0    \label{t-x}
\ee
which is a noncommutative version of the familiar zero curvature condition \cite{Fadd+Takh87}.
We have set $[{\cal U},{\cal V}]_\ast = {\cal U} \ast {\cal V} - {\cal V} \ast {\cal U}$.
The crucial step now is to assume that there is also a (compatible) linear
equation for the parameter $\theta$, i.e.
\be
    \pa_\theta \psi = {\cal W} \ast \psi
\ee
where $\cal W$ is an $N \times N$ matrix of functions (or operators).
This gives rise to further integrability conditions. Using the identity
\be
   \pa_\theta (f \ast g) = (\pa_\theta f) \ast g + f \ast (\pa_\theta g)
   + {1 \over 2} \Big( (\pa_t f) \ast (\pa_x g)-(\pa_x f) \ast (\pa_t g) \Big) \, ,
    \label{diffstar-id}
\ee
we find the following equations:
\be
   \pa_\theta {\cal U} &=& \pa_x {\cal W} + [{\cal W},{\cal U}]_\ast
     + {1 \over 2} [(\pa_x {\cal U}) \ast {\cal V} - (\pa_t {\cal U}) \ast {\cal U}]
    \label{x-th} \\
   \pa_\theta {\cal V} &=& \pa_t {\cal W} + [{\cal W},{\cal V}]_\ast
     + {1 \over 2} [(\pa_x {\cal V}) \ast {\cal V} - (\pa_t {\cal V}) \ast {\cal U}] \; .
    \label{t-th}
\ee
These are zero curvature conditions like (\ref{t-x}), but modified because of
the $\theta$-dependence of the $\ast$-product. It is this modification which
leads to new structures. Because of the explicit dependence of the right hand sides
of (\ref{x-th}) and (\ref{t-th}) on $\theta$, through the use of the $\ast$-product,
these equations are non-autonomous.
\vskip.1cm

Suppose we have achieved in formulating a deformed soliton equation for a variable $u$,
on which ${\cal U}$ and ${\cal V}$ depend, in the form (\ref{t-x}).
If, as a consequence of the specific form of our choices of ${\cal U}$ and ${\cal V}$,
the last two equations reduce to a single (deformation) equation of the form
$\pa_\theta u = F$ where $F$ stands for an expression in $u$ and its $t$- and $x$-derivatives,
but no $\theta$-derivatives, then $\pa_\theta u = F$ is automatically consistent with the
deformed soliton equation. This means that the flows defined by the soliton equation
and the deformation equation commute.
\vskip.1cm

This takes us to another point of view. A characteristic feature of classical soliton
equations is the existence of infinitely many symmetries. A symmetry is given by an
evolution equation which commutes with the soliton equation (see \cite{MJD00}, for
example). It is an evolution equation with respect to a new coordinate, say $\theta$, and
one treats the soliton field, say $u$, now as a function of $x,t,\theta$.
What we are doing is very similar, but differs in so far that we allow for an explicit
dependence of the soliton equation on this new coordinate $\theta$. We restrict this
explicit dependence on $\theta$ to the $\ast$-product, however.
\vskip.1cm

The symmetries of classical soliton equations lead to hierarchies. These are families of
commuting evolution equations. They are easily (Moyal-) deformed to hierarchies of
noncommutative equations. But then we can ask for further symmetries in the sense
explained above. In sections~\ref{sec:hierarchy} and \ref{sec:ncAKNS} we elaborate
this problem for the case of the AKNS hierarchy (see \cite{Ablo+Clark91,FNR83,Newe85},
for example, and the references given there), which admits reductions to the
NLS, KdV and mKdV hierarchies.
It turns out that the noncommutative AKNS hierarchy can be enlarged
to a much bigger hierarchy which includes a new hierarchy of deformation equations.
The corresponding mathematical framework is developed in section~\ref{sec:hierarchy}.
Section~\ref{sec:ncAKNS} then specializes this framework a bit. In particular, it
provides explicit formulae for the first members of the noncommutative versions
of the NLS, KdV and mKdV hierarchies, including their SW deformation hierarchies.
\vskip.1cm

In section~\ref{sec:ncKP} we derive an extension of the noncommutative
version \cite{Hama03b} of the KP hierarchy \cite{Gelf+Dick76,OSTT88,Kupe00}.
Section~\ref{sec:concl} contains some conclusions.

\section{A hierarchy of Lax equations and associated SW deformation equations}
\label{sec:hierarchy}
\setcounter{equation}{0}
Let us consider a family of Lax equations \cite{Wils81,Drin+Soko84}
\be
   \pa_{t_n} {\cal V} = [ {\cal V}^{(n)} , {\cal V} ]_\ast
   \qquad n=0,1,\ldots     \label{Lax-h}
\ee
with\footnote{We are dealing with formal series in $\lambda$,
no convergence is required.}
\be
   {\cal V} = \sum_{k=0}^\infty \lambda^{-k} \, V_k \, , \quad
   {\cal V}^{(n)} = (\lambda^n {\cal V})_{\geq 0}
\ee
where the $V_k$ do not depend on the (formal) parameter $\lambda$ and the
last expression means taking the part of $\lambda^n {\cal V}$ which
corresponds to non-negative powers of $\lambda$, hence
$(\lambda^n {\cal V})_{\geq 0} = \sum_{k=0}^n \lambda^{n-k} \, V_k$. Then
\be
    \bar{{\cal V}}^{(n)} = (\lambda^n {\cal V})_{<0}
                         = \lambda^n {\cal V} - {\cal V}^{(n)}
                         = \sum_{k=1}^\infty \lambda^{-k} \, V_{n+k}
\ee
only contains terms with negative powers of $\lambda$.
An example of the structure (\ref{Lax-h}) (without the deformation)
appeared in Ref.~\cite{FNR83} as a formulation of the AKNS hierarchy.
The $\ast$-product of two functions of the variables $t_0, t_1, \ldots$
is now defined by
\be
   f \ast g = \mathbf{m} \circ e^{\sum_{m,n} \theta_{m,n} P_{m,n}/4} (f\otimes g) \, ,
  \qquad
   P_{m,n} = \pa_{t_m} \otimes \pa_{t_n} - \pa_{t_n} \otimes \pa_{t_m}
   \label{Moyal-h}
\ee
where $\theta_{m,n}=-\theta_{n,m}$ are deformation parameters. Using
\be
   [ {\cal V}^{(n)} , {\cal V} ]_\ast = \sum_{j=1}^\infty \lambda^{-j}
  \sum_{k=0}^{ \min \{ j-1,n \} } [ V_k , V_{n+j-k} ]_\ast \, ,
\ee
(\ref{Lax-h}) turns out to be equivalent to
\be
   \pa_{t_n} V_0 = 0 \, , \quad
   \pa_{t_n} V_j = \sum_{k=0}^{\min\{j-1,n\}} [V_k , V_{n+j-k}]_\ast
   \qquad  j=1,2, \ldots \; .
   \label{Lax-h2}
\ee
We recall an important result of soliton analysis (see
Refs.~\cite{Wils81,FNR83,Drin+Soko84}, for example).
\vskip.2cm

\noindent
{\bf Theorem 1.} The Lax equations (\ref{Lax-h}) imply
\be
   \pa_{t_n} {\cal V}^{(m)} - \pa_{t_m} {\cal V}^{(n)}
 + [ {\cal V}^{(m)} , {\cal V}^{(n)} ]_\ast = 0     \label{0curv-h}
\ee
for all $m,n \geq 0$. The flows (\ref{Lax-h}) commute with each other.
\vskip.1cm

\noindent
{\bf Proof:} We consider two of equations (\ref{Lax-h}) corresponding to
different non-negative integers $n$ and $m$.
Multiplying the first by $\lambda^m$, and decomposing it into non-negative
and negative parts (with respect to powers of $\lambda$) we find
\bez
  \pa_{t_n} {\cal V}^{(m)} + [ {\cal V}^{(m)} , {\cal V}^{(n)} ]_\ast
  + \pa_{t_n} \bar{{\cal V}}^{(m)} - [{\cal V}^{(n)} , \bar{{\cal V}}^{(m)}]_\ast = 0 \; .
\eez
In the same way we obtain
\bez
    \pa_{t_m} {\cal V}^{(n)} + [ {\cal V}^{(n)} , {\cal V}^{(m)} ]_\ast
  + \pa_{t_m} \bar{{\cal V}}^{(n)}
  - [{\cal V}^{(m)} , \bar{{\cal V}}^{(n)}]_\ast = 0 \; .
\eez
Taking the difference of both equations and using
\bez
  0 = [\lambda^n {\cal V}, \lambda^m {\cal V}]_\ast
    = [ {\cal V}^{(n)} , {\cal V}^{(m)} ]_\ast
      + [ {\cal V}^{(n)} , \bar{{\cal V}}^{(m)} ]_\ast
      - [ {\cal V}^{(m)} , \bar{{\cal V}}^{(n)} ]_\ast
      + [ \bar{{\cal V}}^{(n)} , \bar{{\cal V}}^{(m)} ]_\ast
\eez
yields
\bez
    \pa_{t_n} {\cal V}^{(m)} - \pa_{t_m} {\cal V}^{(n)} + [ {\cal V}^{(m)} , {\cal V}^{(n)} ]_\ast
  = \pa_{t_m} \bar{{\cal V}}^{(n)} - \pa_{t_n} \bar{{\cal V}}^{(m)}
    - [\bar{{\cal V}}^{(n)} , \bar{{\cal V}}^{(m)} ]_\ast \; .
\eez
Since both sides of this equation live in different spaces, we conclude that
(\ref{0curv-h}) holds.\footnote{Writing (\ref{Lax-h}) in the equivalent form
$\pa_{t_n} {\cal V} = - [ \bar{{\cal V}}^{(n)} , {\cal V} ]_\ast$,
we see that $\pa_{t_n} \bar{{\cal V}}^{(m)} - \pa_{t_m} \bar{{\cal V}}^{(n)}
- [\bar{{\cal V}}^{(m)} , \bar{{\cal V}}^{(n)} ]_\ast = 0$ is equivalent
to (\ref{0curv-h}).}
Using the Jacobi identity, we find
\bez
   (\pa_{t_n} \pa_{t_m} - \pa_{t_m} \pa_{t_n} ) \, {\cal V}
 = [ \pa_{t_n} {\cal V}^{(m)} - \pa_{t_m} {\cal V}^{(n)} + [ {\cal V}^{(m)} , {\cal V}^{(n)} ]_\ast
   , {\cal V} ]_\ast = 0
\eez
which expresses the commutativity of the flows.
 {  }   \hfill   \rule{5pt}{5pt}
\vskip.2cm

Theorem 1 justifies calling (\ref{Lax-h}) a \emph{hierarchy} of Lax equations.
For fixed $n$, (\ref{Lax-h}) is the compatibility condition of the linear
system\footnote{We assume that $\lambda$ is not only independent of the
coordinates $t_n$, but also does not depend on the deformation parameters.}
\be
   \pa_{t_n} \psi = {\cal V}^{(n)} \ast \psi \, , \quad
   {\cal V} \ast \psi = \lambda \, \psi \; .
\ee
The integrability condition for two members of the linear system, corresponding
to evolution parameters $t_m$ and $t_n$, is (\ref{0curv-h}) which we have shown
to be a consequence of (\ref{Lax-h}).
Following the recipe of the introduction, we extend the family of linear
systems with
\be
   \pa_{\theta_{m,n}} \psi = {\cal W}^{(m,n)} \ast \psi
   \qquad       m,n=0,1,\ldots
\ee
where ${\cal W}^{(m,n)} = - {\cal W}^{(n,m)}$.
The identity (\ref{diffstar-id}) now takes the form
\be
   \pa_{\theta_{m,n}} (f \ast g) = (\pa_{\theta_{m,n}} f) \ast g + f \ast (\pa_{\theta_{m,n}} g)
   + {1 \over 2} \Big( (\pa_{t_m} f) \ast (\pa_{t_n} g)-(\pa_{t_n} f) \ast (\pa_{t_m} g) \Big) \; .
    \label{diffstar-id-h}
\ee
With its help we obtain the additional integrability conditions
\be
 \pa_{t_n} {\cal W}^{(k,l)} - \pa_{\theta_{k,l}} {\cal V}^{(n)}
   &=& [{\cal V}^{(n)} , {\cal W}^{(k,l)} ]_\ast
       + \frac{1}{2} ( \pa_{t_k} {\cal V}^{(n)} \ast {\cal V}^{(l)}
       - \pa_{t_l} {\cal V}^{(n)} \ast {\cal V}^{(k)} )    \label{W_t} \\
 \pa_{\theta_{m,n}} {\cal V}
   &=& [ {\cal W}^{(m,n)} , {\cal V} ]_\ast
       - \frac{1}{2} \Big( \pa_{t_m} {\cal V} \ast {\cal V}^{(n)}
         - \pa_{t_n} {\cal V} \ast {\cal V}^{(m)} \Big)   \label{V_theta-h}
\ee
and
\be
 0 &=& \pa_{\theta_{k,l}} {\cal W}^{(m,n)} - \pa_{\theta_{m,n}} {\cal W}^{(k,l)}
       + [ {\cal W}^{(m,n)} , {\cal W}^{(k,l)} ]_\ast  \nonumber \\
   && + \frac{1}{2} ( \pa_{t_k} {\cal W}^{(m,n)} \ast {\cal V}^{(l)}
      - \pa_{t_l} {\cal W}^{(m,n)} \ast {\cal V}^{(k)}
      - \pa_{t_m} {\cal W}^{(k,l)} \ast {\cal V}^{(n)}
      + \pa_{t_n} {\cal W}^{(k,l)} \ast {\cal V}^{(m)} ) \; . \qquad
      \label{W_theta}
\ee
The above linear system was mainly needed to find the SW deformation equation
(\ref{V_theta-h}). If (\ref{W_t}) holds, it follows that the
Lax flows (\ref{Lax-h}) commute with the flows given by (\ref{V_theta-h}).
Moreover, if (\ref{W_theta}) holds, the deformation flows (\ref{V_theta-h})
commute. After determining ${\cal W}^{(m,n)}$, we show that these conditions
are indeed satisfied as a consequence of (\ref{Lax-h}) and (\ref{V_theta-h}).
\vskip.2cm

$V_0$ does not depend on the coordinates $t_n$ and we assume that it does also
not depend on the deformation parameters $\theta_{m,n}$. As a consequence,
$(\pa_{\theta_{m,n}} {\cal V})_{\geq 0} = 0$, and since
\be
    \Big( (\pa_{t_m} {\cal V}) \ast \bar{{\cal V}}^{(n)} \Big)_{\geq 0}
  = 0
  = \Big((\pa_{t_n} {\cal V}) \ast \bar{{\cal V}}^{(m)} \Big)_{\geq 0} \, ,
\ee
equation (\ref{V_theta-h}) implies
\be
      \Big( [{\cal W}^{(m,n)},{\cal V}]_\ast \Big)_{\geq 0}
  &=& {1 \over 2} \Big( (\pa_{t_m}{\cal V}) \ast {\cal V}^{(n)}
      - (\pa_{t_n}{\cal V}) \ast {\cal V}^{(m)} \Big)_{\geq 0} \nonumber \\
  &=& {1 \over 2} \Big( (\pa_{t_m} \la^n {\cal V}) \ast {\cal V}
      -(\pa_{t_n} \la^m {\cal V}) \ast {\cal V} \Big)_{\geq 0} \nonumber \\
  &=& {1 \over 2} \Big( (\pa_{t_m} {\cal V}^{(n)}) \ast {\cal V}
      -(\pa_{t_n} {\cal V}^{(m)}) \ast {\cal V} \Big)_{\geq 0}  \nonumber \\
  &=& {1 \over 2} \Big( [{\cal V}^{(m)},{\cal V}^{(n)}]_\ast \ast {\cal V}
      \Big)_{\geq 0} \; .
\ee
In the last step we used (\ref{0curv-h}). The last expression is further
evaluated as follows,
\be
     \Big( [{\cal V}^{(m)},{\cal V}^{(n)}]_\ast \ast {\cal V} \Big)_{\geq 0}
 &=& \Big( [{\cal V}^{(m)},\la^n {\cal V} -\bar{{\cal V}}^{(n)}]_\ast \ast {\cal V} \Big)_{\geq 0}
     \nonumber \\
 &=& \Big( [{\cal V}^{(m)} \ast (\la^n {\cal V}),{\cal V}]_\ast
     - [\la^m {\cal V}-\bar{{\cal V}}^{(m)},\bar{{\cal V}}^{(n)}]_\ast \ast {\cal V} \Big)_{\geq 0}
     \nonumber \\
 &=& \Big( [{\cal V}^{(m)} \ast(\la^n {\cal V}),{\cal V}]_\ast
     + [\bar{{\cal V}}^{(n)} \ast (\la^m {\cal V}),{\cal V}]_\ast \Big)_{\geq 0} \nonumber \\
 &=& \Big( -[\bar{{\cal V}}^{(m)} \ast (\la^n {\cal V}),{\cal V}]_\ast
     + [\bar{{\cal V}}^{(n)} \ast (\la^m {\cal V}),{\cal V}]_\ast \Big)_{\geq 0}
\ee
since $[(\la^m {\cal V}) \ast (\la^n {\cal V}),{\cal V}]=0$. Hence
\be
     \Big( [{\cal W}^{(m,n)},{\cal V}]_\ast \Big)_{\geq 0}
 &=& {1 \over 2} \Big( [\bar{{\cal V}}^{(n)} \ast {\cal V}^{(m)}
     - \bar{{\cal V}}^{(m)} \ast {\cal V}^{(n)},{\cal V}]_\ast \Big)_{\geq 0} \; .
\ee
This suggests to choose
\be
  {\cal W}^{(m,n)} = {1 \over 2} \Big( \bar{{\cal V}}^{(n)} \ast {\cal V}^{(m)}
    - \bar{{\cal V}}^{(m)} \ast {\cal V}^{(n)} \Big)_{\geq 0}   \label{Wmn}
\ee
which solves the `non-negative' part of (\ref{V_theta-h}). Inserting the expansions
for the ${\cal V}^{(n)}$ and the $\bar{\cal V}^{(n)}$ in powers of $\lambda$, we find
\be
   \bar{{\cal V}}^{(n)} \ast {\cal V}^{(m)} = \sum_{-m+1}^\infty \lambda^{-j}
   \sum_{k=0}^{ \min\{m-1+j,m\} } V_{m+n+j-k} \ast V_k
\ee
and thus
\be
     {\cal W}^{(m,n)}
 &=& -{1\over2} \sum_{j=0}^{n-1} \lambda^j \sum_{k=1}^{\min\{ n-j,n-m \}} V_{m+k}\ast V_{n-j-k}
     \quad \mbox{if} \quad m < n \; .
\ee
In order to elaborate the SW deformation equations (\ref{V_theta-h}), it is
convenient to insert (\ref{Wmn}) and to use (\ref{Lax-h}) to rewrite them
in the form
\be
   \pa_{\theta_{m,n}} {\cal V}
 = [ {\cal V} , \bar{\cal W}^{(m,n)} ]_\ast
   + {1 \over 2} (\bar{{\cal V}}^{(n)} \ast \pa_{t_m}{\cal V}
   - \bar{{\cal V}}^{(m)} \ast \pa_{t_n}{\cal V})   \label{SW-h}
\ee
where
\be
      \bar{\cal W}^{(m,n)}
  &=& {1 \over 2} \Big( \bar{{\cal V}}^{(n)} \ast {\cal V}^{(m)}
       -\bar{{\cal V}}^{(m)} \ast {\cal V}^{(n)} \Big)_{<0}  \nonumber \\
  &=& - \frac{1}{2} \sum_{j=1}^\infty \lambda^{-j} \sum_{k=1}^{n-m}
      V_{n+j-k} \ast V_{m+k}  \qquad \mbox{if} \quad m<n \; .
\ee
Of course, we can use (\ref{Lax-h}) to eliminate the $t$-derivatives in the last
two terms of (\ref{SW-h}). As the $\lambda^{-1}$ part of (\ref{SW-h}) we find
\be
   \pa_{\theta_{m,n}} V_1
 = - \frac{1}{2} [ V_0 , \sum_{k=m+1}^n V_{m+n+1-k} \ast V_k ]_\ast
   \qquad \mbox{if} \quad m<n \; .    \label{V1_theta}
\ee
\vskip.2cm

\noindent
{\bf Theorem 2.} The integrability condition (\ref{W_t}), with ${\cal W}^{(m,n)}$
defined in (\ref{Wmn}), holds as a consequence of (\ref{Lax-h}) and (\ref{V_theta-h}).
The flows given by (\ref{Lax-h}) commute with the flows (\ref{V_theta-h}).
\vskip.1cm
\noindent
{\bf Proof:}
Multiplication of (\ref{SW-h}) with $\lambda^k$ and restriction to the non-negative
part leads to
\bez
     \pa_{\theta_{m,n}} {\cal V}^{(k)}
  = \Big( [ {\cal V}^{(k)}, \bar{\cal W}^{(m,n)} ]_\ast
    + {1 \over 2} ( \bar{{\cal V}}^{(n)} \ast \pa_{t_m} {\cal V}^{(k)}
     - \bar{{\cal V}}^{(m)} \ast \pa_{t_n} {\cal V}^{(k)} ) \Big)_{\geq 0} \; .
\eez
Application of $\pa_{t_k}$ to (\ref{Wmn}) yields
\bez
   \pa_{t_k} {\cal W}^{(m,n)}
 &=& \frac{1}{2} (
        \pa_{t_m} {\cal V}^{(k)} \ast {\cal V}^{(n)}
      - \pa_{t_n} {\cal V}^{(k)} \ast {\cal V}^{(m)} )
      + [ {\cal V}^{(k)} , {\cal W}^{(m,n)} ]_\ast   \\
 & &  + \Big( [ {\cal V}^{(k)} , \bar{\cal W}^{(m,n)} ]_\ast
      + \frac{1}{2} ( \bar{\cal V}^{(n)} \ast \pa_{t_m} {\cal V}^{(k)}
      - \bar{\cal V}^{(m)} \ast \pa_{t_n} {\cal V}^{(k)} ) \Big)_{\geq 0}  \; .
\eez
Here we used (\ref{0curv-h}) in the form
\bez
    \pa_{t_k}{\cal V}^{(n)}
  = \pa_{t_n}{\cal V}^{(k)} + [{\cal V}^{(k)},{\cal V}^{(n)}]_\ast
\eez
which with the help of (\ref{Lax-h}) implies
\bez
     \pa_{t_k}\bar{{\cal V}}^{(n)}
  = -\pa_{t_n}{\cal V}^{(k)} + [{\cal V}^{(k)},\bar{{\cal V}}^{(n)}]_\ast \; .
\eez
Now we obtain
\bez
     \pa_{t_k}{\cal W}^{(m,n)} - \pa_{\theta_{m,n}}{\cal V}^{(k)}
  = [{\cal V}^{(k)},{\cal W}^{(m,n)}]_\ast
     + {1 \over 2} (\pa_{t_m} {\cal V}^{(k)} \ast {\cal V}^{(n)}
     - \pa_{t_n}{\cal V}^{(k)} \ast {\cal V}^{(m)}) \; .
\eez
This is the integrability condition (\ref{W_t}) which in fact implies
the commutativity of the $t$- and $\theta$-flows.
{  }   \hfill   \rule{5pt}{5pt}
\vskip.2cm

\noindent
{\bf Theorem 3.} The integrability condition (\ref{W_theta}), with ${\cal W}^{(m,n)}$
defined in (\ref{Wmn}), holds as a consequence of (\ref{Lax-h}) and (\ref{V_theta-h}).
The $\theta$-flows given by (\ref{V_theta-h}) commute for all $m,n \geq 0$.
\vskip.1cm
\noindent
{\bf Proof:} This is a straightforward but tedious calculation.
Let us write $A^{(m,n,k,l)}$ for the expression on the right hand side of
(\ref{W_theta}). First we derive
\bez
 & & \pa_{\theta_{k,l}} {\cal W}^{(m,n)}  \\
 &=& {1 \over 2}\Big( \pa_{t_m}{\cal W}^{(k,l)} \ast {\cal V}^{(n)}
     - \pa_{t_n}{\cal W}^{(k,l)} \ast {\cal V}^{(m)}
     + \bar{{\cal V}}^{(n)} \ast \pa_{t_m}{\cal W}^{(k,l)}
     - \bar{{\cal V}}^{(m)} \ast \pa_{t_n} {\cal W}^{(k,l)} \\
 & & -\pa_{t_k}\tilde{{\cal W}}^{(m,n)} \ast {\cal V}^{(l)}
     + \pa_{t_l}\tilde{{\cal W}}^{(m,n)} \ast {\cal V}^{(k)}
     -2 \, [\tilde{{\cal W}}^{(m,n)},{\cal W}^{(k,l)}]_\ast  \\
 & & + {1 \over 2}(\pa_{t_k}\bar{{\cal V}}^{(n)} \ast \pa_{t_m}{\cal V}^{(l)}
     - \pa_{t_l} \bar{{\cal V}}^{(n)} \ast \pa_{t_m}{\cal V}^{(k)}
     - \pa_{t_k}\bar{{\cal V}}^{(m)} \ast \pa_{t_n}{\cal V}^{(l)}
     + \pa_{t_l}\bar{{\cal V}}^{(m)} \ast \pa_{t_n}{\cal V}^{(k)}) \Big)_{\geq 0}
\eez
where $\tilde{{\cal W}}^{(m,n)} = {\cal W}^{(m,n)} + \bar{{\cal W}}^{(m,n)}$.
Here we used the expression for $\pa_{\theta_{k,l}}{\cal V}^{(n)}$ obtained in the
proof of Theorem 2 and a corresponding expression for $\pa_{\theta_{k,l}} \bar{{\cal V}}^{(n)}$.
Then we applied (\ref{0curv-h}) and (\ref{W_t}) (which holds according to Theorem 2)
several times. With the help of the above expression, we find
\bez
      A^{(m,n,k,l)}
 &=& {1 \over 2} \Big( -2 \, [\tilde{{\cal W}}^{(m,n)},\tilde{{\cal W}}^{(k,l)}]_\ast
     + \bar{{\cal V}}^{(n)} \ast \pa_{t_m}\tilde{{\cal W}}^{(k,l)}
     - \bar{{\cal V}}^{(m)} \ast \pa_{t_n}\tilde{{\cal W}}^{(k,l)} \\
 & & -\bar{{\cal V}}^{(l)} \ast \pa_{t_k}\tilde{{\cal W}}^{(m,n)}
     + \bar{{\cal V}}^{(k)} \ast \pa_{t_l}\tilde{{\cal W}}^{(m,n)}
     + \pa_{t_m}\tilde{{\cal W}}^{(k,l)} \ast {\cal V}^{(n)}
     - \pa_{t_n}\tilde{{\cal W}}^{(k,l)} \ast {\cal V}^{(m)} \\
 & & -\pa_{t_k}\tilde{{\cal W}}^{(m,n)} \ast {\cal V}^{(l)}
     + \pa_{t_l}\tilde{{\cal W}}^{(m,n)} \ast {\cal V}^{(k)}
     + {1 \over 2} (\pa_{t_k}\bar{{\cal V}}^{(n)} \ast \pa_{t_m}{\cal V}^{(l)}
     - \pa_{t_l}\bar{{\cal V}}^{(n)} \ast \pa_{t_m}{\cal V}^{(k)} \\
 & & -\pa_{t_k}\bar{{\cal V}}^{(m)} \ast \pa_{t_n}{\cal V}^{(l)}
     + \pa_{t_l}\bar{{\cal V}}^{(m)} \ast \pa_{t_n}{\cal V}^{(k)}
     -\pa_{t_m}\bar{{\cal V}}^{(l)} \ast \pa_{t_k}{\cal V}^{(n)}
     + \pa_{t_n}\bar{{\cal V}}^{(l)} \ast \pa_{t_k}{\cal V}^{(m)} \\
 & & + \pa_{t_m}\bar{{\cal V}}^{(k)} \ast \pa_{t_l}{\cal V}^{(n)}
     -\pa_{t_n}\bar{{\cal V}}^{(k)} \ast \pa_{t_l}{\cal V}^{(m)}) \Big)_{\geq 0} \; .
\eez
Inserting $\tilde{{\cal W}}^{(m,n)} = {1 \over 2} (\bar{{\cal V}}^{(n)} \ast {\cal V}^{(m)}
 - \bar{{\cal V}}^{(m)} \ast {\cal V}^{(n)})$
and using (\ref{0curv-h}) in various forms, a lengthy calculation leads to $A^{(m,n,k,l)}=0$,
so that (\ref{W_theta}) is indeed satisfied. This condition implies the commutativity
of the deformation flows.
{  }   \hfill   \rule{5pt}{5pt}
\vskip.2cm

We have thus shown that (\ref{V_theta-h}) defines a hierarchy of deformation
equations which extends the noncommutative hierarchy (\ref{Lax-h}) of Lax equations
to a larger hierarchy.

\section{The ncAKNS hierarchy and its reductions}
\label{sec:ncAKNS}
\setcounter{equation}{0}
Let us look at the compatibility condition of the second member ($n=1$) of the
Lax hierarchy (\ref{Lax-h}) with the $n$-th member ($n > 1$).
Setting $t_1 = x$ and using the notation used in the introduction, we are
dealing with
\be
   {\cal U} = {\cal V}^{(1)} \, , \qquad {\cal V} = {\cal V}^{(n)} \; .
\ee
Furthermore, we set
\be
   V_0 = H \, , \quad V_1 = U
\ee
where $H$ is (in accordance with the first of equations (\ref{Lax-h2}))
a \emph{constant} $N \times N$ matrix, which splits the
algebra $\mbox{Mat}_N$ of $N \times N$ matrices in such a way that
$\mbox{Mat}_N = \mbox{ker}(\mbox{ad} H) \oplus \mbox{Im}(\mbox{ad} H)$
(see also Ref.~\citen{Wils81,Drin+Soko84}).
Every matrix then has a unique decomposition
$M = M^{(+)} + M^{(-)}$ where $M^{(+)} \in \mbox{ker}(\mbox{ad} H)$ and
$M^{(-)} \in \mbox{Im}(\mbox{ad} H)$. For the corresponding parts of
$V_k$ we write
\be
    A_k = V_k^{(-)} \, , \qquad B_k = V_k^{(+)} \; .
\ee
In the following we use lower indices $x,t_n$ to indicate derivatives
of functions or matrices with respect to these variables. For example,
$U_x = \pa_x U$.

Inserting our ansatz in (\ref{t-x}), which is a special case of (\ref{0curv-h}),
we obtain
\be
  U_x &=& [H , V_2]   \label{U_x} \\
  V_{k,x} &=& [U,V_k]_\ast + [H , V_{k+1}]
  \quad \qquad  k = 2,\ldots,n-1 \, , \label{V_x} \\
  U_{t_n} &=& V_{n,x} - [U,V_n]_\ast  \; .
\ee
Considering the family of equations with $n \geq 0$, we may write the last
equation as
\be
    U_{t_n} = [H,V_{n+1}]_\ast
\ee
(see also Ref.~\citen{Wils81}). The above equations are precisely those which
we obtain from the Lax hierarchy (\ref{Lax-h2}).\footnote{See also Lemma (2.4)
in Ref.~\citen{Wils81}.}
Equation (\ref{U_x}) suggests to choose $U \in \mbox{Im}(\mbox{ad} H)$, i.e.
\be
     U = U^{(-)} \; .
\ee
The above system of equations then decomposes into
\be
   U_x &=& [ H , A_2 ] \\
   A_{k,x} &=& [U,B_k]_\ast + [H,A_{k+1}]   \qquad k=2,\ldots,n-1  \\
   B_{k,x} &=& [U,A_k]_\ast  \qquad k=2,\ldots,n-1  \label{B_kx} \\
   B_{n,x} &=& [U,A_n]_\ast \\
   U_{t_n} &=& [H , A_{n+1}]_\ast \; .
\ee
The first equation determines $A_2$. Then $B_2$ can be obtained from the
third equation\footnote{Introducing a formal inverse of $\pa_x$, as common in
soliton analysis, we can always solve for the $B_k$. In the case under
consideration, we checked up to $k=7$ that there is in fact a \emph{local}
solution $B_k$, so that the right hand side of (\ref{B_kx}) can be written
as a total $x$-derivative of products of $U$ and its $x$-derivatives.}
and the second then yields $A_3$. By iteration we find
the $A_k$ and $B_k$. The last equation above is a deformed evolution equation.
\vskip.2cm

In the following, we specialize to
\be
    H = \frac{1}{2 \alpha} \, \sigma  \quad \mbox{with} \quad \sigma^2 = I
\ee
where $\alpha \neq 0$ is a constant and $I$ denotes the $N \times N$ unit matrix.
Algebraically, we are then dealing with a deformation of the AKNS hierarchy
\cite{Ablo+Clark91,FNR83,Newe85} (which was originally expressed in terms
of $2 \times 2$-matrices, see below). Then
\be
   A_k = \alpha \, ( -A_{k-1,x} + [U,B_{k-1}]_\ast ) \, \sigma \, , \quad
   B_{k,x} = [U,A_k]_\ast  \qquad k = 2,3, \ldots
\ee
and
\be
    \alpha \, U_{t_n} = - A_{n+1} \, \sigma
    \qquad n=0,1,\ldots \; .                \label{ncAKNS}
\ee
\vskip.2cm

Besides $A_0 = 0$ and $A_1 = U$, for the first values of $n$ we find
\be
   A_2 &=& - \alpha \, U_x \, \sigma  \\
   A_3 &=& - \alpha^2 \, \Big( - U_{xx} + 2 \, U^{\ast 3} \Big)   \\
   A_4 &=& - \alpha^3 \, \Big( U_{xxx} - 3 \, (U^{\ast 2} \ast U_x
           + U_x \ast U^{\ast 2}) \Big) \, \sigma  \\
   A_5 &=& - \alpha^4 \, \Big( - U_{xxxx} + 4 \, U^{\ast 2} \ast U_{xx}
           + 4 \, U_{xx} \ast U^{\ast 2} + 2 \, U \ast U_{xx} \ast U           \nonumber \\
       && + 6 \, U_x \ast U \ast U_x + 2 \, U \ast U_x{}^{\ast 2} + 2 \, U_x{}^{\ast 2} \ast U
          - 6 \, U^{\ast 5} \Big)  \\
   A_6 &=& - \alpha^5 \, \Big( U_{xxxxx} -5 \, ( U^{\ast 2} \ast U_{xxx}
           + U_{xxx} \ast U^{\ast 2} ) - 10 \, ( U_x \ast U \ast U_{xx} + U_{xx} \ast U \ast U_x )
                   \nonumber \\
       & & - 5 \, ( U \ast U_x \ast U_{xx} + U \ast U_{xx} \ast U_x
           + U_x \ast U_{xx} \ast U + U_{xx} \ast U_x \ast U ) - 10 \, U_x{}^{\ast 3}
                   \nonumber \\
       && + 10 \, ( U_x \ast U^{\ast 4} + U^{\ast 4} \ast U_x
         + U^{\ast 2} \ast U_x \ast U^{\ast 2} ) \Big) \, \sigma   \qquad \\
  A_7 &=& - \alpha^6 \Big( - U_{xxxxxx} + 2 \, U \ast U_{xxxx} \ast U
          + 6 \, ( U^{\ast 2} \ast U_{xxxx} + U_{xxxx} \ast U^{\ast 2} ) \nonumber \\
      & & + 9 \, ( U \ast U_x \ast U_{xxx} + U_{xxx} \ast U_x \ast U )
          + 4 \, ( U \ast U_{xxx} \ast U_x + U_x \ast U_{xxx} \ast U ) \nonumber \\
      & & + 15 \, ( U_x \ast U \ast U_{xxx} + U_{xxx} \ast U \ast U_x )
          + 11 \, ( U \ast U_{xx}{}^{\ast 2} + U_{xx}{}^{\ast 2} \ast U )
          + 20 \, U_{xx} \ast U \ast U_{xx}   \nonumber \\
      & & + 20 \, U_x \ast U_{xx} \ast U_x
          - 15 \, ( U^{\ast 4} \ast U_{xx} + U_{xx} \ast U^{\ast 4} )
          + 25 \, ( U_{xx} \ast U_x{}^{\ast 2} + U_x{}^{\ast 2} \ast U_{xx} ) \nonumber \\
      & & - 10 \, ( U^{\ast 3} \ast U_{xx} \ast U + 2 \, U^{\ast 2} \ast U_{xx} \ast U^{\ast 2}
                  + U \ast U_{xx} \ast U^{\ast 3}  )
          - 10 \, ( U^{\ast 3} \ast U_x{}^{\ast 2} + U_x{}^{\ast 2} \ast U^{\ast 3} ) \nonumber \\
      & & - 15 \, ( U^{\ast 2} \ast U_x{}^{\ast 2} \ast U + U \ast U_x{}^{\ast 2} \ast U^{\ast 2}  )
          - 25 \, ( U^{\ast 2} \ast U_x \ast U \ast U_x + U_x \ast U \ast U_x \ast U^{\ast 2} )
          \nonumber \\
      & & - 5 \, ( U \ast U_x \ast U^{\ast 2} \ast U_x + 2 \, U \ast U_x \ast U \ast U_x \ast U
                 + U_x \ast U^{\ast 2} \ast U_x \ast U   )  \nonumber \\
      & & - 20 \, U_x \ast U^{\ast 3} \ast U_x + 20 \, U^{\ast 7} \Big)   \\
  A_8 &=& - 7 \alpha^7 \, \Big(
         U_{xxxxxxx}/7 - U_{xxxxx} \ast U^{\ast 2} - U^{\ast 2} \ast U_{xxxxx}
         - U \ast U_{xxxx} \ast U_x - U_x \ast U_{xxxx} \ast U  \nonumber \\
   &&      - 2 \, ( U_{xxxx} \ast U_x \ast U + U \ast U_x \ast U_{xxxx} )
           - 3 \, ( U_{xxxx} \ast U \ast U_x + U_x \ast U \ast U_{xxxx} )
           \nonumber \\
   && - 5 \, ( U_{xxx} \ast U \ast U_{xx} + U_{xx} \ast U \ast U_{xxx} )
      - 3 \, ( U_{xxx} \ast U_{xx} \ast U + U \ast U_{xx} \ast U_{xxx} )
           \nonumber \\
   && - 2 \, ( U_{xx} \ast U_{xxx} \ast U + U \ast U_{xxx} \ast U_{xx} )
      - 4 \, U_x \ast U_{xxx} \ast U_x - 7 \, ( U_{xxx} \ast U_x{}^{\ast 2}
              + U_x{}^{\ast 2} \ast U_{xxx} )  \nonumber \\
   && + 4 \, U^{\ast 2} \ast U_{xxx} \ast U^{\ast 2}
      + 3 \, ( U_{xxx} \ast U^{\ast 4} + U^{\ast 4} \ast U_{xxx} )
      - 8 \, ( U_{xx}{}^{\ast 2} \ast U_x + U_x \ast U_{xx}{}^{\ast 2} )
         \nonumber \\
   && - 10 \, U_{xx} \ast U_x \ast U_{xx}
      + 4 \, ( U_{xx} \ast U_x \ast U^{\ast 3} + U^{\ast 3} \ast U_x \ast U_{xx} )
        \nonumber \\
   && + 4 \, ( U^{\ast 2} \ast U_{xx} \ast U_x \ast U + U \ast U_x \ast U_{xx} \ast U^{\ast 2}
           + U_x \ast U_{xx} \ast U^{\ast 3} + U^{\ast 3} \ast U_{xx} \ast U_x )
           \nonumber \\
   && + 8 \, ( U_x \ast U \ast U_{xx} \ast U^{\ast 2} + U^{\ast 2} \ast U_{xx} \ast U \ast U_x )
      + 7 \, ( U_{xx} \ast U \ast U_x \ast U^{\ast 2} + U^{\ast 2} \ast U_x \ast U \ast U_{xx} )
         \nonumber \\
   && + 5 \, ( U_{xx} \ast U^{\ast 3} \ast U_x + U_x \ast U^{\ast 3} \ast U_{xx} )
      + 3 \, ( U_x \ast U^{\ast 2} \ast U_{xx} \ast U + U \ast U_{xx} \ast U^{\ast 2} \ast U_x
        \nonumber \\
   &&      + U^{\ast 2} \ast U_x \ast U_{xx} \ast U + U \ast U_{xx} \ast U_x \ast U^{\ast 2} )
      + 2 \, ( U_{xx} \ast U^{\ast 2} \ast U_x \ast U + U \ast U_x \ast U^{\ast 2} \ast U_{xx} )
             \nonumber \\
   && + 9 \, ( U_x{}^{\ast 3} \ast U^{\ast 2} + U^{\ast 2} \ast U_x{}^{\ast 3} )
         + 6 \, ( U \ast U_x{}^{\ast 2} \ast U \ast U_x + U_x \ast U \ast U_x{}^{\ast 2} \ast U )
           \nonumber \\
   &&    + 10 \, U_x \ast U \ast U_x \ast U \ast U_x
         + 5 \, ( U_x{}^{\ast 2} \ast U^{\ast 2} \ast U_x + U_x \ast U^{\ast 2} \ast U_x{}^{\ast 2} )
           \nonumber \\
   &&    + 4 \, ( U_x{}^{\ast 2} \ast U \ast U_x \ast U + U \ast U_x \ast U \ast U_x{}^{\ast 2} )
         + 2 \, U \ast U_x{}^{\ast 3} \ast U
         \nonumber \\
   && - 5 \, ( U^{\ast 6} \ast U_x + U_x \ast U^{\ast 6} + U^{\ast 4} \ast U_x \ast U^{\ast 2}
              + U^{\ast 2} \ast U_x \ast U^{\ast 4} ) \Big) \, \sigma
\ee
using the notation $U^{\ast 2} = U \ast U$. Since $\sigma$ is a constant matrix, the
cyclic property of the trace requires $\mbox{tr} (A_k \sigma) = 0$, and thus in particular
$\mbox{tr} (U \sigma) = 0$.
Neglecting constants of integration, besides $B_0 = \sigma/2$ and $B_1 = 0$ we find
\be
   B_2 &=& - \alpha \, U^{\ast 2} \, \sigma  \\
   B_3 &=& \alpha^2 \, [U,U_x]_\ast  \\
   B_4 &=& - \alpha^3 \, \Big( U \ast U_{xx} + U_{xx} \ast U - U_x{}^{\ast2}
           - 3 U^{\ast4} \Big) \, \sigma  \\
   B_5 &=& - \alpha^4 \, \Big( [U_{xxx} , U]_\ast - [U_{xx}, U_x]_\ast
           + 4 \, U^{\ast 2} \ast [U, U_x]_\ast + 4 \, [U, U_x]_\ast \ast U^{\ast 2}
           \nonumber \\
       & & + 2 \, U \ast [U,U_x]_\ast \ast U \Big)  \\
   B_6 &=& - \alpha^5 \, \Big( U_{xxxx} \ast U + U \ast U_{xxxx}
           - U_{xxx} \ast U_x - U_x \ast U_{xxx} + U_{xx}{}^{\ast 2} \nonumber \\
       & & - 5 \, ( U^{\ast 3} \ast U_{xx} + U^{\ast 2} \ast U_{xx} \ast U
               + U \ast U_{xx} \ast U^{\ast 2} + U_{xx}*U^{\ast 3} ) \nonumber \\
       & & + 5 \, ( U_x \ast U^{\ast 2} \ast U_x - U \ast U_x \ast U \ast U_x
           - U_x \ast U \ast U_x \ast U - U \ast U_x{}^{\ast 2} \ast U ) \nonumber \\
       & & + 10 \, U^{\ast 6} \Big) \, \sigma   \\
   B_7 &=& - \alpha^6 \, \Big( [ U_{xxxxx},U]_\ast + [U_x,U_{xxxx}]_\ast
          + [U_{xxx},U_{xx}]_\ast + 6 \, [ U^{\ast 3},U_{xxx}]_\ast
          + 4 \, U \ast [U_{xxx},U]_\ast \ast U   \nonumber \\
       && + 3 \, ( U^{\ast 2} \ast U_x \ast U_{xx} - U_{xx} \ast U_x \ast U^{\ast 2} )
          + 9 \, ( U \ast U_x \ast U \ast U_{xx} - U_{xx} \ast U \ast U_x \ast U )
             \nonumber \\
       && + 6 \, ( U_{xx} \ast U^{\ast 2} \ast U_x - U_x \ast U^{\ast 2} \ast U_{xx} )
          + 8 \, ( U^{\ast 2} \ast U_{xx} \ast U_x - U_x \ast U_{xx} \ast U^{\ast 2} )
            \nonumber \\
       && + U \ast [U_{xx},U_x]_\ast \ast U
          + 11 \, ( U \ast U_{xx} \ast U \ast U_x - U_x \ast U \ast U_{xx} \ast U )
          + 13 \, [ U,U_x{}^{\ast 3}]_\ast   \nonumber \\
       && + 3 \, ( U_x \ast U \ast U_x{}^{\ast 2} - U_x{}^{\ast 2} \ast U \ast U_x )
          + 15 \, [U_x,U^{\ast 5}]_\ast
          + 5 \, ( U^{\ast 4} \ast U_x \ast U - U \ast U_x \ast U^{\ast 4} )
             \nonumber \\
       && + 10 \, U^{\ast 2} \ast [U_x,U]_\ast \ast U^{\ast 2} \Big) \; .
\ee

Let us elaborate the SW deformation equations which are given by (\ref{V1_theta}).
We restrict our considerations to the deformations between the second ($n=1$) and
the remaining ncAKNS equations ($n>1$) and write
\be
     \theta_n = \theta_{1,n}  \qquad  n = 0,2, \ldots \; .
\ee
We obtain
\be
    U_{\theta_n}
  = \frac{1}{2 \alpha} \Big( \sum_{k=2}^n V_{n+2-k} \ast V_k \Big)^{(-)} \sigma
  = \frac{1}{2 \alpha} \sum_{k=2}^n ( A_k \ast B_{n+2-k} + B_k \ast A_{n+2-k} )
    \, \sigma             \label{ncAKNS-SW}
\ee
which determines SW deformation equations for the members of the hierarchy.
The first equations, corresponding to $n=0,2,3,4,5,6$, are
\be
   U_{\theta_0} &=& 0  \\
   U_{\theta_2} &=& {\alpha \over 2} [ U_x , U^{\ast 2} ]_\ast \, \sigma  \\
   U_{\theta_3} &=& {\alpha^2 \over 2} \Big( [U^{\ast 2} , U_{xx} ]_\ast + [ U_x{}^{\ast 2} , U]_\ast
                    \Big)   \\
   U_{\theta_4} &=& {\alpha^3 \over 2} \Big( [ U_{xxx}, U^{\ast 2}]_\ast + [U_x , U_{xx}]_\ast \ast U
                    + U \ast [U_x , U_{xx}]_\ast + 4 \, [U^{\ast 4} , U_x]_\ast
                    \nonumber \\
                & & + 2 \, U \ast [U^{\ast 2} , \ast U_x] \ast U \Big) \, \sigma
                    \label{AKNS_SW2} \\
   U_{\theta_5} &=& {\alpha^4 \over 2} \Big( [U^{\ast 2} , U_{xxxx}]_\ast
                    + [U_x \ast U_{xxx} + U_{xxx} \ast U_x - U_{xx}{}^{\ast 2}, U]_\ast
                    + 5 \, [U_{xx} , U^{\ast 4} ]_\ast  \nonumber \\
                & & + 5 \, ( U \ast [U_x{}^{\ast 2} , U]_\ast \ast U
                    + [U_x \ast U , U_x \ast U^{\ast 2}]_\ast
                    + [U \ast U_x , U^{\ast 2} \ast U_x]_\ast \Big)
                    \label{AKNS_SW3}   \\
   U_{\theta_6} &=& {\alpha^5 \over 2} \Big( [U_{xxxxx}, U^{\ast 2}]_\ast
        + U \ast [U_x , U_{xxxx}]_\ast + [U_x , U_{xxxx}]_\ast \ast U
        + U \ast [U_{xxx}, U_{xx}]_\ast    \nonumber \\
   &&   + [U_{xxx} , U_{xx}]_\ast \ast U
        + 6 \, [U^{\ast 4} , U_{xxx}]_\ast + 2 \, U \ast [ U^{\ast 2} , U_{xxx}]_\ast \ast U
               \nonumber \\
   &&   + 9 \, ( U^{\ast 2} \ast U_x \ast U \ast U_{xx}
               + U^{\ast 2} \ast U_{xx} \ast U_x \ast U
               - U \ast U_x \ast U_{xx} \ast U^{\ast 2}
               - U_{xx} \ast U \ast U_x \ast U^{\ast 2} )  \nonumber \\
   &&   + 6 \, ( U \ast U_{xx} \ast U^{\ast 2} \ast U_x
               + U_{xx} \ast U^{\ast 2} \ast U_x \ast U
               - U \ast U_x \ast U^{\ast 2} \ast U_{xx}
               - U_x \ast U^{\ast 2} \ast U_{xx} \ast U ) \nonumber \\
   &&   + 3 \, ( U^{\ast 3} \ast U_x \ast U_{xx} - U_{xx} \ast U_x \ast U^{\ast 3} )
        + 8 \, ( U^{\ast 3} \ast U_{xx} \ast U_x - U_x \ast U_{xx} \ast U^{\ast 3} )
           \nonumber \\
   &&   + 2 \, ( U^{\ast 2} \ast U_x \ast U_{xx} \ast U
               - U \ast U_{xx} \ast U_x \ast U^{\ast 2}
               + U \ast U_{xx} \ast U \ast U_x \ast U
           \nonumber \\
   &&          - U \ast U_x \ast U \ast U_{xx} \ast U )
        + 11 \, ( U^{\ast 2} \ast U_{xx} \ast U \ast U_x
                - U_x \ast U \ast U_{xx} \ast U^{\ast 2} )
        + 13 \, [U^{\ast 2}, U_x{}^{\ast 3}]_\ast
           \nonumber \\
   &&   + 3 \, [U \ast U_x \ast U , U_x{}^{\ast 2}]_\ast
        + 3 \, [U_x, U \ast U_x{}^{\ast 2} \ast U]_\ast
        + 5 \, ( U^{\ast 2} \ast [U_x,U^{\ast 2}]_\ast \ast U^{\ast 2}     \nonumber \\
   &&         + 2 \, U \ast [U_x , U^{\ast 4}]_\ast \ast U
              + 3 \, [U_x,U^{\ast 6}]_\ast ) \Big) \, \sigma   \\
   U_{\theta_7} &=& {\alpha^6 \over 2} \Big( [U^{\ast 2},U_{xxxxxx}]_\ast
      + [U_x \ast U_{xxxxx} + U_{xxxxx} \ast U_x , U]_\ast
      + [U , U_{xx} \ast U_{xxxx} + U_{xxxx} \ast U_{xx}]_\ast  \nonumber \\
   && + 7 \, [U_{xxxx}, U^{\ast 4}]_\ast + [U_{xxx}{}^{\ast 2},U]_\ast
      + 7 \, \Big( U_{xxx} \ast U_x \ast U^{\ast 3} - U^{\ast 3} \ast U_x \ast U_{xxx}
            \nonumber \\
   && + 2 \, ( U_{xxx} \ast U \ast U_x \ast U^{\ast 2} - U^{\ast 2} \ast U_x \ast U \ast U_{xxx} )
      + U \ast U_x \ast U^{\ast 2} \ast U_{xxx}
                \nonumber \\
   && - U_{xxx} \ast U^{\ast 2} \ast U_x \ast U
      + U \ast U_{xxx} \ast U_x \ast U^{\ast 2} - U^{\ast 2} \ast U_x \ast U_{xxx} \ast U
          \nonumber \\
   && + U \ast U_{xxx} \ast U^{\ast 2} \ast U_x - U_x \ast U^{\ast 2} \ast U_{xxx} \ast U
      + 2 \, ( U_x \ast U \ast U_{xxx} \ast U^{\ast 2}
      \nonumber \\
   && - U^{\ast 2} \ast U_{xxx} \ast U \ast U_x
      + U \ast U_x \ast U_{xxx} \ast U^{\ast 2} - U^{\ast 2} \ast U_{xxx} \ast U_x \ast U )
        \nonumber  \\
   && + 2 \, [ U_{xx}{}^{\ast 2}, U^{\ast 3}]_\ast
             + 3 \, [U_{xx} \ast U \ast U_{xx} , U^{\ast 2}]_\ast
             + U \ast [U_{xx}{}^{\ast 2}, U]_\ast \ast U
             + [U_{xx} \ast U^{\ast 2} \ast U_{xx}, U]_\ast   \nonumber \\
   && + 4 \, ( U_{xx} \ast U_x{}^{\ast 2} \ast U^{\ast 2}
             - U^{\ast 2} \ast U_x{}^{\ast 2} \ast U_{xx} )
              + U \ast U_x{}^{\ast 2} \ast U \ast U_{xx}
             - U_{xx} \ast U \ast U_x{}^{\ast 2} \ast U
             \nonumber \\
   && + 2 \, [U_x \ast U_{xx} \ast U_x , U^{\ast 2} ]_\ast
      + 2 \, ( U \ast U_x \ast U \ast U_{xx} \ast U_x - U_x \ast U_{xx} \ast U \ast U_x \ast U )
        \nonumber \\
   && + U \ast U_x{}^{\ast 2} \ast U_{xx} \ast U - U \ast U_{xx} \ast U_x{}^{\ast 2} \ast U
      + U \ast U_{xx} \ast U \ast U_x{}^{\ast 2} - U_x{}^{\ast 2} \ast U \ast U_{xx} \ast U
      \nonumber \\
   && + 3 \, ( U_x{}^{\ast 2} \ast U_{xx} \ast U^{\ast 2} - U^{\ast 2} \ast U_{xx} \ast U_x{}^{\ast 2} )
      + 2 \, ( U \ast U_x \ast U_{xx} \ast U \ast U_x
      \nonumber \\
   && - U_x \ast U \ast U_{xx} \ast U_x \ast U )
      + 3 \, [ U^{\ast 6}, U_{xx}]_\ast + U^{\ast 2} \ast [ U^{\ast 2}, U_{xx}]_\ast U^{\ast 2}
      + 3 \, [ U, U_x{}^{\ast 4}]_\ast   \nonumber \\
   && + [ U^{\ast 5}, U_x{}^{\ast 2}]_\ast
      + 4 \, [U^{\ast 4} , U_x \ast U \ast U_x ]_\ast + [ U_x \ast U^{\ast 2} \ast U_x,U^{\ast 3}]_\ast
      + 2 \, [ U^{\ast 2}, U_x \ast U^{\ast 3} \ast U_x]_\ast  \nonumber \\
   && + U^{\ast 2} \ast [ U_x{}^{\ast 2} , U ]_\ast \ast U^{\ast 2}
      + 3 \, [ U_x \ast U^{\ast 4} \ast U_x , U]_\ast
      + 3 \, U \ast  [ U^{\ast 3} , U_x{}^{\ast 2}]_\ast \ast U   \nonumber \\
   && + 3 \, U \ast [ U , U_x \ast U^{\ast 2} \ast U_x]_\ast \ast U \Big) \Big)  \; .
\ee
For each $n= 2,3, \ldots$, (\ref{ncAKNS}) determines a deformed evolution equation and
(\ref{ncAKNS-SW}) yields the associated SW deformation equation.
Using the computer algebra program FORM \cite{Verm00,Verm02} we checked the commutativity
of the ncAKNS flows and their SW deformation flows in the cases $n=2,\ldots,7$.
In section~\ref{sec:hierarchy} we provided corresponding general proofs.
\vskip.2cm

Choosing $2 \times 2$-matrices (hence $N=2$) and for $\sigma$ the Pauli matrix
\be
    \sigma_3 = \mbox{diag}(1,-1) \, ,
\ee
we are dealing with the original AKNS hierarchy \cite{Ablo+Clark91,FNR83,Newe85}.
We have extended it to a `noncommutative space' by introducing a Moyal-deformation
with respect to the coordinates involved and derived corresponding SW deformation
equations. Several integrable models including the KdV, mKdV and NLS equation
are obtained as reductions of the AKNS hierarchy \cite{Ablo+Clark91,FNR83,Newe85}.
Corresponding deformed soliton equations and hierarchies result from the above
deformed AKNS hierarchy.
In the following we list some examples and derive from (\ref{ncAKNS-SW}) the
corresponding SW deformation equations.
\vskip.1cm
\noindent
(1) For even $n$, let us choose $\alpha = -i$ and  \\
\be
   U = \left( \begin{array}{cc} 0 & q \\ \pm \bar{q} & 0 \end{array} \right)
\ee
where $\bar{q}$ denotes the complex conjugate of the function $q$. We assume
$\bar{\theta} = \theta$ in this case. From (\ref{ncAKNS}) we obtain the hierarchy
\be
   i \, q_{t_2} &=& - q_{xx} \pm 2 q \ast \bar{q} \ast q    \label{ncNLS} \\
   i \, q_{t_4} &=& -\Big(
    q_{xxxx} \mp 4 q \ast \bar{q} \ast q_{xx} \mp 2 q \ast \bar{q}_{xx} \ast q
   \mp 4 q_{xx} \ast \bar{q} \ast q \mp 2 q \ast \bar{q}_x \ast q_x \nonumber \\
   && \mp 6 q_x \ast \bar{q} \ast q_x \mp 2 q_x \ast \bar{q}_x \ast q
    + 6 q \ast \bar{q} \ast q \ast \bar{q} \ast q \Big)  \\
   &\vdots&    \nonumber
\ee
where the first equation is the ncNLS equation \cite{DMH01ncNLS,DMH01FK},
the deformation of the NLS equation. The corresponding SW deformation hierarchy
begins with
\be
  q_{\theta_2} &=& \pm {i \over 2} ( q_x \ast \bar{q} \ast q - q \ast \bar{q} \ast q_x )
     \label{SW-NLS} \\
  q_{\theta_4} &=& \pm {i \over 2} [ q_{xxx} \ast \bar{q} \ast q - q \ast \bar{q} \ast q_{xxx}
              + (q_x \ast \bar{q}_{xx} - q_{xx} \ast \bar{q}_x) \ast q
              + q \ast ( \bar{q}_x \ast q_{xx} - \bar{q}_{xx} \ast q_x) ] \nonumber \\
             & & + i \, [ 2 (q \ast \bar{q})^{\ast 2} \ast q_x
                - 2 q_x \ast (\bar{q} \ast q)^{\ast 2}
                + q \ast (\bar{q} \ast q \ast \bar{q}_x - \bar{q}_x \ast q \ast \bar{q}) \ast q ]
                \; .
\ee
For $n=6$ already rather lengthy expressions arise which we omit here. They can be quite
easily generated from our general formulae. Equation (\ref{SW-NLS}) already appeared in
\cite{DMH01ncNLS}.
\vskip.1cm
\noindent
(2) For odd $n$, $\alpha = 1$, and
\be
  U = \left( \begin{array}{cc} 0 & u \\ 1 & 0 \end{array} \right) \, ,
\ee
we obtain the hierarchy\footnote{The fact that the right hand sides
can be written as total $x$-derivatives, so that these equations are conservation
laws with common conserved density $u$, is a special feature of this reduction.}
\be
   u_{t_3} &=& \Big( u_{xx} - 3 \, u^{\ast 2} \Big)_x   \\
   u_{t_5} &=& \Big( u_{xxxx} - 5 \, ( u_{xx} \ast u + u \ast u_{xx}
                   + u_x{}^{\ast 2} ) + 10 \, u^{\ast 3} \Big)_x   \\
   u_{t_7} &=& \Big( u_{xxxxxx} - 7 \, ( u_{xxxx} \ast u + u \ast u_{xxxx} )
              - 14 \, ( u_{xxx} \ast u_x + u_x \ast u_{xxx} )
               \nonumber \\
           && - 21 \, u_{xx}{}^2
              + 21 \, ( u_{xx} \ast u^{\ast 2} + u^{\ast 2} \ast u_{xx} )
              + 28 \, u \ast u_{xx} \ast u  \nonumber \\
           && + 14 \, ( u_x \ast u \ast u_x + 2 \ast u \ast u_x{}^{\ast 2}
                       + 2 \, u_x{}^{\ast 2} \ast u )
              + 35 \, u^{\ast 4}  \Big)_x   \\
           & \vdots &    \nonumber
\ee
The first equation is the ncKdV equation \cite{DMH00ncKdV,Pani01},
the deformation of the KdV equation.
The corresponding SW deformation hierarchy starts with
\be
  u_{\theta_3} &=& \frac{1}{2} [u,u_{xx}]_\ast = \frac{1}{2} \Big( [u,u_x ]_\ast \Big)_x   \\
  u_{\theta_5} &=& \frac{1}{2} \Big( [u,u_{xxx}]_\ast
                   + [u_{xx},u_x]_\ast + 5 \, [u_x,u^{\ast2}]_\ast \Big)_x  \\
  u_{\theta_7} &=& \frac{1}{2} \Big( [u,u_{xxxxx}]_\ast
                  + [u_{xxxx},u_x]_\ast + [u_{xx},u_{xxx}]_\ast
                  + 7 \, ( [u_{xxx},u^{\ast2}]_\ast + [(u_x{}^{\ast 2})_x,u]_\ast  \nonumber \\
               & &+ [ u_x ,  u \ast u_{xx} + u_{xx} \ast u ]_\ast + u \ast [u,u_x]_\ast \ast u )
                  + 21 \, [u^{\ast 3},u_x]_\ast \Big)_x   \; .
\ee
The first equation was obtained in a different way in Ref.~\citen{DMH00ncKdV}
where it has been used to calculate nc-corrections to the two-soliton
solution of the KdV equation.
\vskip.1cm
\noindent
(2) We restrict again to odd $n$, but this time we choose
\be
   U = \left(\begin{array}{cc} 0 & v \\ v & 0 \end{array} \right)
\ee
which leads to the ncmKdV hierarchy
\be
   v_{t_3} &=&
    v_{xxx} - 3 \, (v \ast v \ast v_x + v_x \ast v \ast v)   \\
   v_{t_5} &=& - \Big( 5 \, ( v_{xxx} \ast v^{\ast 2} + v^{\ast 2} \ast v_{xxx} )
       + 5 \, (v_{xx} \ast v_x \ast v + v \ast v_x \ast v_{xx}
        \nonumber \\
    && + v_x \ast v_{xx} \ast v + v \ast v_{xx} \ast v_x
       + 2 \, v_{xx} \ast v \ast v_x + 2 \, v_x \ast v \ast v_{xx} )
       + 10 \, v_x{}^{\ast 3}   \nonumber \\
    && - 10 \, ( v_x \ast v^{\ast 4} + v^{\ast 2} \ast v_x \ast v^{\ast 2}
       + v^{\ast 4} \ast v_x ) - v^{\ast 5} \Big)  \\
   &\vdots&  \nonumber
\ee
The corresponding SW deformation hierarchy starts with
\be
   v_{\theta_3} &=& {1 \over 2} ([v \ast v,v_{xx}]_\ast - [v,v_x \ast v_x]_\ast)  \\
   v_{\theta_5} &=& {1 \over 2} ( [v^{\ast 2} , v_{xxxx}]_\ast
         + [ v_x \ast v_{xxx} + v_{xxx} \ast v_x , v]_\ast)
         + [ v , v_{xx}{}^{\ast 2} ]_\ast
         + 5 \, [ v_{xx} , v^{\ast 4} ]_\ast \nonumber \\
     &&  + 5 \, v \ast [v_x{}^{\ast 2} , v ]_\ast \ast v
         + 5 \, [ v \ast v_x , v^{\ast 2} \ast v_x ]_\ast
         + 5 \, [ v_x \ast v , v_x \ast v^{\ast 2} ]_\ast ) \; .
\ee
The $n=1$ equations were obtained in a different way in Ref.~\citen{DMH00ncKdV}.
\vskip.2cm

\noindent
{\bf Remarks.} \\
(1) For some fixed $n > 1$ we may set $\theta_{1,n} = 0$ with the effect that
the $\ast$-product does not contain derivatives with respect to the `time' $t_n$.
In this case we still have an ordinary evolution equation and the deformation
only involves the `spatial' coordinate $x$ and the remaining $t_m$ where $m \neq n$.
Interpreting $t_n$ as a physical time coordinate would mean that, with
$\theta_{1,n} \neq 0$, we were dealing with a `space-time' deformation, which
is rather speculative, though such deformations have been discussed in the context
of string theories. This deformation turns the classical evolution equation into
one which is non-local in time.
The interpretation of any of the parameters $t_n$ appearing in the above hierarchy
as a `time' in a physical sense is not obligatory, however.  \\
(2) More generally, we may consider deformations between each pair of levels $m,n \geq 0$,
allowing $\theta_{m,n} \neq 0$. With the assumptions made in this section,
the deformation equations (\ref{V1_theta}) take the form
\be
   \pa_{\theta_{(m,n)}} U = \frac{1}{2 \alpha} \sum_{k=1}^{n-m} ( A_{n+1-k} \ast B_{m+k}
    + B_{n+1-k} \ast A_{m+k} ) \, \sigma
\ee
which have to be evaluated by inserting the above expressions for the $A_k$ and $B_k$.
For example, in the ncKdV case we find
\be
      u_{\theta_{3,5}}
 &=& {1 \over 2} \Big( [u_{xx},u_{xxx}]_\ast
     + 3 \, [u_{xxx},u^{\ast2}]_\ast
     + 2 \, [(u^{\ast2})_x , u_{xx}]_\ast
     + 2 \, ( u \ast u_x \ast u_{xx} - u_{xx} \ast u_x \ast u) \nonumber \\
 & & + 12 \, u \ast [u,u_x]_\ast \ast u
     + 6 \, [u^{\ast3},u_x]_\ast \Big)_x  \; .
\ee
(3) Without deformation, the AKNS hierarchy has common conserved densities
given by $\mbox{tr}( V_k \, \sigma ) = \mbox{tr}( B_k \, \sigma )$ \cite{Wils81}.
Due to the non-commutativity of the $\ast$-product, these are in general no longer
expressions for the conserved densities of the deformed hierarchy.

\section{The extended ncKP hierarchy}
\label{sec:ncKP}
\setcounter{equation}{0}
Let
\be
   L = \pa + \sum_{j=1}^\infty u_{j+1} \, \pa^{-j} \, , \quad
   L^{(n)} = (L^n)_{\geq 0} \, , \quad
   \bar{L}^{(n)} = (L^n)_{<0} = L^n-L^{(n)}
\ee
where we simply write $L^n$ instead of $L^{\ast n}$.
The non-negative (negative) part of a formal series is now understood in
the sense of non-negative (negative) powers of $\pa=\pa_x$.
Using the derivation property of $\pa$ and
\be
   \pa^{-1} u_k = u_k \, \pa^{-1} - u_{k,x} \, \pa^{-2} + u_{k,xx} \, \pa^{-3}
   - u_{k,xxx} \, \pa^{-4} + \ldots
\ee
we obtain
\be
   L^{(1)} &=& \pa   \\
   L^{(2)} &=& \pa^2 + 2 \, u_2  \\
   L^{(3)} &=& \pa^3 + 3 \, u_2 \, \pa + 3 ( u_{2,x} + u_3 )  \\
   L^{(4)} &=& \pa^4 + 4 \, u_2 \, \pa^2 + ( 6 \, u_{2,x} + 4 \, u_3 ) \, \pa
               + 4 \, u_{2,xx} + 6 \, u_2^{\ast 2} + 6 \, u_{3,x} + 4 \, u_4 \; .
\ee
The \emph{ncKP hierarchy} (see also Ref.~\citen{Hama03b}) is the set of equations
\be
    \pa_{t_n} L = [L^{(n)},L]_\ast    \qquad   n = 1,2, \ldots  \label{ncKP}
\ee
where $t_1 = x$. This is a deformation of the classical KP hierarchy \cite{Gelf+Dick76,OSTT88,Kupe00},
using the $\ast$-product given by (\ref{Moyal-h}). Elaborating these equations for
$n=2,3$, we find
\be
   \pa_{t_2} u_2 &=& ( u_{2,x} + 2 \, u_3 )_x \label{ncKP22} \\
   \pa_{t_2} u_3 &=& u_{3,xx} + 2 \, u_{4,x} + 2 \, u_2 \ast u_{2,x} + 2 \, [ u_2 , u_3]_\ast
                      \label{ncKP23} \\
   \pa_{t_2} u_4 &=& 2 \, u_{5,x} + u_{4,xx} + 4 \, u_3 \ast u_{2,x}
                   - 2 \, u_2 \ast u_{2,xx} + 2 \, [ u_2 , u_4]_\ast \\
   \pa_{t_2} u_5 &=& 2 \, u_{6,x} + u_{5,xx} + 2 \, u_2 \ast u_{2,xxx}
                   - 6 \, u_3 \ast u_{2,xx} + 6 \, u_4 \ast u_{2,x} + 2 \, [ u_2 , u_5]_\ast \\
                 &\vdots& \nonumber  \\
   \pa_{t_3} u_2 &=& ( u_{2,xx} + 3 \, u_2^{\ast 2} + 3 \, u_{3,x} + 3 \, u_4 )_x
                      \label{ncKP32} \\
   \pa_{t_3} u_3 &=& u_{3,xxx} + 3 \, u_{4,xx} + 6 \, u_2 \ast u_{3,x}
            + 3 \, u_{2,x} \ast u_3 + 3 \, u_3 \ast u_{2,x}   \nonumber \\
                 & & + 3 \, u_{5,x} + 3 \, [ u_2 , u_4 ]_\ast \\
   \pa_{t_3} u_4 &=& u_{4,xxx} + 3 \, u_{5,xx} - 3 \, u_2 \ast u_{3,xx}
                     - 3 \, u_3 \ast u_{2,xx} + 6 \, u_4 \ast u_{2,x}  \nonumber \\
                 & & + 3 \, ( u_2 \ast u_4 )_x + 3 \, u_{6,x}
                     + 3 \, [ u_3 , u_4]_\ast + 3 \, [ u_2 , u_5 ]_\ast \\
   \pa_{t_3} u_5 &=& 3 \, u_{7,x} + u_{5,xxx} + 3 \, u_{6,xx} + 3 \, u_3 \ast u_{2,xxx}
                     + 3 \, u_2 \ast u_{3,xxx} - 9 \, u_3 \ast u_{3,xx}
                      \nonumber \\
                 & & - 9 \, u_4 \ast u_{2,xx} + 3 \, ( u_2 \ast u_5 )_x
                     + 9 \, u_5 \ast u_{2,x}  + 9 \, u_4 \ast u_{3,x}
                     + 3 \, [ u_2 , u_6 ]_\ast + 3 \, [ u_3 , u_5 ]_\ast \qquad \\
                 &\vdots&  \nonumber
\ee
(see also Ref.~\cite{Hama03b}).
Eliminating $u_3$ and $u_4$ from (\ref{ncKP22}), (\ref{ncKP23}) and (\ref{ncKP32}), one
recovers a deformed version \cite{Pani01,Hama03b} of the KP equation,
\be
   \Big( 4 \, u_{2,t} - u_{2,xxx} - 6 (u_2^{\ast 2})_x
   + 6 \, [u_2 , \int u_{2,y} \, dx \, ]_\ast \Big)_x - 3 \, u_{2,yy} = 0
\ee
where $t_2 = y$ and $t_3 = t$.

The ncKP equations imply $\pa_{t_n} L^m = [L^{(n)},L^m]_\ast$ which, using
\be
   0 = [L^n,L^m]_\ast = [L^{(n)},L^{(m)}]_\ast + [L^{(n)},\bar{L}^{(m)}]_\ast
       + [\bar{L}^{(n)},L^{(m)}]_\ast + [\bar{L}^{(n)},\bar{L}^{(m)}]_\ast \, ,
\ee
leads to
\be
   \pa_{t_n} L^{(m)} - \pa_{t_m} L^{(n)} + [L^{(m)},L^{(n)}]_\ast = 0
   \label{ncKP-cfc}
\ee
and
\be
  \pa_{t_n} \bar{L}^{(m)} - \pa_{t_m} \bar{L}^{(n)} - [\bar{L}^{(m)},\bar{L}^{(n)}]_\ast = 0
\ee
which is equivalent to (\ref{ncKP-cfc}).
As a consequence of this equation and the Jacobi identity, the flows given
by (\ref{ncKP}) indeed commute. Using (\ref{ncKP}), (\ref{ncKP-cfc}) can also be written as
\be
   \pa_{t_n} \bar{L}^{(m)} + \pa_{t_m} L^{(n)} + [\bar{L}^{(m)},L^{(n)}]_\ast = 0 \; .
\ee

The $n$-th member of the ncKP hierarchy arises as integrability condition
of the linear system
\be
   L \ast \psi = \lambda \psi \, , \qquad
   \pa_{t_n} \psi = L^{(n)} \ast \psi
\ee
assuming $\pa_{t_n} \lambda =0$. The integrability condition for two different
members of the linear system, with coordinates $t_n$ and $t_m$, respectively,
is precisely (\ref{ncKP-cfc}) which we have seen to be satisfied as a consequence
of (\ref{ncKP}). Following the recipe of the introduction, we
extend the above linear system by imposing deformation equations of the form
\be
   \pa_{\theta_{mn}}\psi = W^{(m,n)} \ast \psi
\ee
assuming $\pa_{\theta_{mn}} \lambda =0$. This results in new integrability conditions.
We find
\be
   \pa_{\theta_{mn}} L
 = [W^{(m,n)},L]_\ast + {1 \over 2} \Big( \pa_{t_n} L\ast L^{(m)} - \pa_{t_m}L \ast L^{(n)} \Big)
   \label{ncKP-SW}
\ee
and, by induction and using (\ref{diffstar-id-h}),
\be
   \pa_{\theta_{mn}} L^r
 = [W^{(m,n)},L^r]_\ast + {1 \over 2} \Big( (\pa_{t_n} L^r) \ast L^{(m)}
   - (\pa_{t_m} L^r)\ast L^{(n)} \Big) \; .
\ee
Further integrability conditions are
\be
  \pa_{t_r} W^{(m,n)} - \pa_{\theta_{mn}} L^{(r)} + [W^{(m,n)},L^{(r)}]_\ast =
  {1 \over 2} \Big( \pa_{t_m} L^{(r)} \ast L^{(n)} - \pa_{t_n} L^{(r)} \ast L^{(m)} \Big)
\ee
and
\be
 0 &=& \pa_{\theta_{rs}} W^{(m,n)} - \pa_{\theta_{mn}} W^{(r,s)} + [W^{(m,n)},W^{(r,s)}]_\ast
       \nonumber \\
   & & - {1 \over 2} (\pa_{t_m} W^{(r,s)} \ast L^{(n)} - \pa_{t_n} W^{(r,s)} \ast L^{(m)}
       - \pa_{t_r} W^{(m,n)} \ast L^{(s)} + \pa_{t_s} W^{(m,n)} \ast L^{(r)}) \; . \qquad
\ee
Since $(\pa_{\theta_{mn}}L)_{\geq 0} = 0$, (\ref{ncKP-SW}) implies
\be
      \Big( [W^{(m,n)},L]_\ast \Big)_{\geq 0}
  &=& {1 \over 2} \Big( \pa_{t_m} L \ast L^{(n)} - \pa_{t_n} L \ast L^{(m)} \Big)_{\geq 0} \nonumber \\
  &=& {1 \over 2} \Big( [L^{(m)},L]_\ast \ast L^n - [L^{(n)},L]_\ast \ast L^m \Big)_{\geq 0} \nonumber \\
  &=& {1 \over 2} \Big( [L^{(m)} \ast L^n,L]_\ast - [L^{(n)} \ast L^m,L]_\ast \Big)_{\geq 0} \nonumber \\
  &=& {1 \over 2} \Big( [ - \bar{L}^{(m)} \ast L^n + \bar{L}^{(n)} \ast L^m , L]_\ast \Big)_{\geq 0} \nonumber \\
  &=& {1 \over 2} \Big( [ - \bar{L}^{(m)} \ast L^{(n)} + \bar{L}^{(n)} \ast L^{(m)} , L]_\ast \Big)_{\geq 0}
\ee
which suggests to set
\be
  W^{(m,n)} = {1 \over 2} \Big( \bar{L}^{(n)} \ast L^{(m)} - \bar{L}^{(m)} \ast L^{(n)}\Big)_{\geq 0} \; .
  \label{ncKP-Wmn}
\ee
This expression is completely analogous to the corresponding ncAKNS expression (\ref{Wmn}).
Introducing
\be
  \bar{W}^{(m,n)} = {1 \over 2} \Big( \bar{L}^{(n)} \ast L^{(m)} - \bar{L}^{(m)} \ast L^{(n)} \Big)_{<0}
\ee
we also obtain
\be
   \pa_{\theta_{mn}} L^r
 = [L^r,\bar{W}^{(m,n)}]_\ast + {1 \over 2} \Big( \bar{L}^{(n)} \ast \pa_{t_m} L^r
   - \bar{L}^{(m)} \ast \pa_{t_n} L^r \Big) \; . \label{ncKP-SW2}
\ee
As a consequence of this equation and (\ref{ncKP}), the above integrability conditions
are satisfied and it follows that the flows given by (\ref{ncKP-SW}), with $W^{(m,n)}$
defined in (\ref{ncKP-Wmn}), commute with each other and also commute with the flows
given by (\ref{ncKP}). The corresponding calculations are rather tedious, however.
We thus obtained an extension of the ncKP hierarchy.
\vskip.2cm

In particular, from (\ref{ncKP-SW2}) we obtain
\be
   \pa_{\theta_{mn}} u_2
 = {1 \over 2} \pa_x \Big( \bar{L}^{(n)} \ast L^{(m)} - \bar{L}^{(m)} \ast L^{(n)} \Big)_{-1}
\ee
so that, for example,
\be
   \pa_{\theta_{1,2}} u_2
 = {1 \over 2} \Big( u_4 + u_{3,x} - u_2^{\ast 2} \Big)_x
 =  {1 \over 6} \, \Big( u_{2,t} - u_{2,xxx} - 6 \, (u_2^{\ast 2})_x \Big)  \; .
\ee
Using FORM \cite{Verm00,Verm02} we checked that the flow determined by this equation
indeed commutes with the ncKP flow.

\section{Conclusions}
\label{sec:concl}
\setcounter{equation}{0}
We have shown that the noncommutative AKNS and KP hierarchies can be extended to
larger hierarchies which contain corresponding SW deformation equations.
Of course, our procedure to generate deformation equations for which the
flow commutes with the flow of the deformed equation one starts with (and thus
a symmetry of the latter equation) can be applied to other deformed
integrable models and corresponding hierarchies. One can say that, as a
consequence of the explicit dependence on deformation parameters, deformed
soliton equations have additional symmetries.
We have demonstrated that there is much more structure behind \emph{deformed}
soliton equations than has been revealed up to now.
\vskip.2cm

As in the case of the ncKdV equation treated in Ref.~\citen{DMH00ncKdV},
SW deformation equations can be used to construct solutions of the
corresponding nc-soliton equation from solutions of its classical version.
\vskip.2cm

Conserved densities have already been obtained for several noncommutative
versions of soliton equations, including the ncNLS and ncKdV equation (see
Refs.~\citen{DMH00ncKdV,DMH01ncNLS,DMH01FK,Hama03b}, in particular).
It turned out that these cannot be expressed in terms of the $\ast$-product
only. Another `generalized $\ast$-product' was needed, which is commutative,
but non-associative. It has been introduced in Ref.~\citen{Stra97} and
frequently used in recent work on noncommutative field theories
(see Refs.~\citen{Garo00,Liu01}, for example).
Because of technical problems associated with the generalized $\ast$-product,
the question of \emph{common} conserved densities of extended noncommutative
soliton hierarchies has not found an answer yet. A better understanding of
the properties of the generalized $\ast$-product seems to be required.

\end{document}